\documentclass[a4paper,12pt]{article}
\usepackage{latexsym,amssymb}
\usepackage{graphicx}
\usepackage{cite}
\usepackage[arrow,matrix,curve]{xy}
\usepackage{amsmath}

\global\arraycolsep=2pt 
\input{epsf} 

\def\s[#1,#2]{[#1\stackrel{{\displaystyle\star}}{,}#2]}

 1

\usepackage{amsthm}

\newcommand{\eq}{\begin{equation}}
\newcommand{\eqa}{\begin{eqnarray}}
\newcommand{\en}{\end{equation}}
\newcommand{\ena}{\end{eqnarray}}
\newcommand{\enn}{\nonumber \end{equation}}

\def\sk{\vskip .4cm}
\def\noi{\noindent}

\def\epsi{{\varepsilon}}
\def\st {\star}
\def\f{{\rm{f}\,}}
\def\of{{\overline{{\rm{f}}\,}}}

\def\D/h{\widehat{\fmslash D}}

\def\om{\omega}
\def\Om{\Omega}
\def\al{\alpha}

\def\be{\beta}
\def\ga{\gamma}
\def\Ga{\Gamma}

\def\5bar{{\overline 5}}

\def\RR{{\mathcal R}}

\def\R{{R}}
\def\oR{{\overline{\R}}}

\def\UU{{U}\Xi}

\def\FF{\mathcal F}

\def\varepsi{\varepsilon}

\def\s'O{\stackrel{_{{\displaystyle\st \footnotesize '}}}{_{^{^{\displaystyle\otimes}}}}}

\def\AAs{{A_\st }}

\def\Xis{{\Xi_\st }}
\def\Oms{\Omega_\st}

\def\ll{{\mathcal L}}
\def\D{\Delta}
\def\1s{{1_\st }}
\def\3s{{3_\st }}
\def\2s{{2_\st }}
\def\ef1{{1_\FF}}
\def\ef2{{3_\FF}}
\def\ef3{{2_\FF}}

\def\TT{{\mathcal T}}

\def\Diff{\mbox{\it Diff}\,}

\def\le{\langle}
\def\re{\rangle}
\def\nn{\nonumber}

\def\dd{{\triangledown}}
\def\dds{\triangledown^\st}

\def\rr{\mathsf{R}}
\def\ric{\mathsf{Ric}}
\def\tr{\mathsf{T}}

\def\ots{\otimes_\st}
\def\res{\re_\st}

\def\x{\mathsf{x}}
\def\y{\mathsf{y}}
  
\def\r4{\mathbb{R}^4}  
\def\UP{U(iso(3,1))}
\def\Diff{\mbox{\sl Diff}}
\numberwithin{equation}{section}

\begin{document}

\begin{titlepage}
\rightline{DISTA-UPO/05}
\sk\sk\sk\sk
\begin{center}
{\bf\LARGE{Noncommutative Symmetries and Gravity$\,^{^{_{_\st}}}$}}\\[.5em] 
\vskip 2.5em

{{\bf Paolo Aschieri}}

\vskip 1.5em
The Erwin Schr\"odinger International Institute for Mathematical Physics,\\
Bolzmanngasse 9, A-1090 Vienna, Austria\\[1em]

Dipartimento di Scienze e Tecnologie Avanzate\\
Universit\' a del Piemonte Orientale, and INFN\\
Via Bellini 25/G 15100 Alessandria, Italy\\[1em]
\end{center}

\sk
\sk
\sk
\centerline{\bf Abstract}
\sk
\normalsize{Spacetime geometry is twisted (deformed) 
into noncommutative spacetime geometry,  where
functions and tensors are now star-multiplied.
Consistently, spacetime diffeomorhisms are twisted into noncommutative
diffeomorphisms.  Their deformed Lie
algebra structure and that of infinitesimal Poincar\'e transformations 
is defined and explicitly constructed. 

This allows to construct a noncommutative theory of gravity. }
\sk\sk
\sk\sk

\noi This article is based on common work with Christian Blohmann, Marija 
Dimitrijevi\'c, Frank Meyer, Peter Schupp and Julius Wess.

\sk\sk
\noi

{\footnotesize{\noi PACS: 02.40.Gh, 02.20.Uw, 04.20.-q,  11.10.Nx, 04.60.-m.~~~ 
}}
\sk\sk
\noi{$^\st\,$Extended version of a minicourse held at the workshop ``Noncommutative Geometry in Field and String Theories", Corfu Summer
Institute on EPP, September 2005, Corfu, Greece.}  
\sk
\noi {\footnotesize{E-mail: aschieri@theorie.physik.uni-muenchen.de}}\\

\sk\sk


\end{titlepage}\vskip.2cm

\newpage

\section{Introduction}




The study of gravity on noncommutative spacetime, where spacetime uncertainty
relations and nonlocal effects naturally arise, is an interesting arena
for the study of spacetime at Planck length, 
where quantum gravity effects are non-negligible.
This line of thought has been pursued since the early days of
quantum mechanics
\cite{Pauli:1985},
and more recently in  \cite{Madore:1992} -\cite{GR2}, (see also the
recent review \cite{Szabo}).

This work is based on \cite{G1} and \cite{GR2}, where we
study the algebra of diffeomorphisms on noncommutative spacetime and 
a noncommutative gravity theory covariant under diffeomorphisms. In
\cite{G1} we study the case of constant noncommutativity, 
$x^\mu\st x^\nu-x^\nu\st x^\mu=\theta^{\mu\nu}$, while in \cite{GR2} 
we consider a quite general class of noncommutative manifolds 
obtained via Drinfeld
twists; there generally the commutator 
$x^\mu\st x^\nu-x^\nu\st x^\mu$ is a nonconstant function. 
We here emphasize the twist approach to noncommutative
gravity, and we develop the notion of noncommutative Lie algebra,
including a detailed study of the noncommutative Poincar\'e Lie
algebra. For pedagogical reasons we treat just the case of
constant noncommutativity 
(and we assume that commutative spacetime as a manifold is $\r4$).

\sk
In Section 2 we introduce the twist 
$\FF=e^{-{i\over 2}\theta^{\mu\nu}{\partial_\mu}\otimes{\partial_\nu}}$.
The general notion of twist is well
known \cite{Drinfeld1, Drinfeld3}. 
Multiparametric twists appear in \cite{Reshetikhin}.
In the context of deformed Poincar\'e group and Minkowski space geometry 
twists have been studied in \cite{ACPoincare}, \cite{Kulish-Mudrov},
\cite{Lukierski} (multiparametric deformations),
and in \cite{Oeckl:2000eg}-\cite{Watts} 
(Moyal-Weyl deformations).

Given a twist $\FF$ we  state the general principle that allows to construct
noncommutative products by composing commutative products with the
twist $\FF$. In this way we obtain the algebrae of noncommutative
functions, tensorfields, exterior forms and diffeomorphisms. 
Noncommutative 
diffeomorphisms are then shown to naturally act on tensorfields and
forms. We study in detail the notion of infinitesimal 
diffeomorphism, and the corresponding notion of deformed Lie algebra.

In Section 3 we present the example of the Poincar\'e symmetry, give 
explicitly the infinitesimal generators and their deformed Lie bracket, 
and explain the geometric origin of the latter.
The generators and the bracket differ from the ones usually considered
in the literature.

In Section 4 we use the noncommutative differential geometry
formalism introduced in Section 2 and develop the notion of covariant
derivative, and of torsion, curvature and Ricci curvature tensors. 
In Section 5 a metric on noncommutative space is introduced. The corresponding
unique torsion free metric compatible connection is used to construct
the Ricci tensor and obtain the Eintein equations for gravity on
noncommutative spacetime.

In the Appendix we show that the algebra of differential operators is
not a Hopf algebra, and we relate it to 
the Hopf algebra of infinitesimal diffeomorphisms.

\section{Deformation by twists}

A quite general procedure in order to construct noncommutative spaces and
noncommutative field theories is that of  a twist. 
The ingredients are:\\

I) a Lie algebra $g$.\\

II) an action of the Lie algebra
   on the space one wants to deform. \\

III) a twist element $\FF$, constructed with the generators 
   of the Lie algebra $g$. \\

Concerning III), a twist element $\FF$ is an invertible element in 
   $Ug\otimes Ug$, where
   $Ug$ is the universal enveloping algebra of $g$. $Ug$ is a Hopf algebra,
   in particular there is a linear map, called coproduct 
   \eq \Delta:Ug\rightarrow Ug\otimes Ug~.\en
For every Lie algebra element $t\in g$ we have 
\eq 
\Delta(t)=t\otimes 1 +1\otimes t\, .
\en
The coproduct $\D$ is extended to all $Ug$ by defining 
$$\D(tt'):=\D(t)\D(t')=tt'\otimes 1+t\otimes t' +t'\otimes t+ 1\otimes tt'$$
and more generally $\D(tt'\ldots t'')=\D(t)\D(t')\ldots\D(t'')$.
A main property $\FF$ has to satisfy is the cocycle condition
\eq
\label{propF1}
(\FF\otimes 1)(\Delta\otimes id)\FF=(1\otimes \FF)(id\otimes \Delta)\FF\,~.
\en
\sk
If $g$ is the Lie algebra of vectorfields on spacetime $M=\r4$, 
or simply the subalgebra spanned by  the 
commuting vectorfields ${\partial/\partial x^\mu}$, we can consider the twist 
\eq\label{MWTW}
\FF=e^{-{i\over 2}\theta^{\mu\nu}{\partial\over \partial x^\mu}
\otimes{\partial\over \partial x^\nu}}~,
\en 
where $\theta^{\mu\nu}$ is a real antisymmetric matrix of 
dimensionful constants. 
We consider $\theta^{\mu\nu}$  fundamental physical constants,
like the velocity of light $c$, or like $\hbar$. The symmetries of our physical 
system will leave $\theta^{\mu\nu},\,c$ and $\hbar$ invariant.
The inverse of $\FF$ is
$$\FF^{-1}=e^{{i\over 2}\theta^{\mu\nu}{\partial\over \partial x^\mu}
\otimes{\partial\over \partial x^\nu}}~.$$
This twist satisfies condition (\ref{propF1}) because the Lie algebra 
of partial derivatives is abelian. 
\sk
The star-product between functions can be obtained from the usual 
pointwise product via the action of the twist operator,
namely,
\eq\label{starprodf}
f\st g:=\mu\circ \FF^{-1}(f\otimes g)~,
\en
where $\mu$ is the usual pointwise product between functions, 
$\mu(f\otimes g)=fg$. 

We shall frequently use the notation (sum over $\al$ understood)
\eq\label{Fff}
\FF=\f^\al\otimes\f_\al~~~,~~~~\FF^{-1}=\of^\al\otimes\of_\al~,
\en
so that
\eq\label{fhfg}
f\st g:=\of^\al(f)\of_\al(g)~.
\en
Explicitly we have
\eq
\FF^{-1}=e^{{i\over 2}\theta^{\mu\nu}{\partial\over \partial x^\mu}
\otimes{\partial\over \partial x^\nu}}
=\sum {1\over{n!}}\left( i\over 2\right)^n\theta^{\mu_1\nu_1}\ldots\theta^{\mu_n\nu_n}
\partial_{\mu_1}\ldots\partial_{\mu_n}\otimes
{}\partial_{\nu_1}\ldots\partial_{\nu_n}=\of^\al\otimes\of_\al~,\label{faexp}
\en
so that $\al$ is a multi-index.
We also introduce the universal $\RR$-matrix
\eq
\RR:=\FF_{21}\FF^{-1}~\label{defUR}
\en
where by definition $\FF_{21}=\f_\al\otimes \f^\al$.
In the sequel we use the notation 
\eq
\RR=\R^\al\otimes\R_\al~~~,~~~~~~\RR^{-1}=\oR^\al\otimes\oR_\al~.
\en
In the present case we simply have $\RR=\FF^{-2}$ but for
more general twists this is no more the case.
The $\RR$-matrix measures the noncommutativity of the $\star$-product.
Indeed it is easy to see that 
\eq\label{Rpermutation}
h\st g=
\oR^\al(g)\st\oR_\al(h)~.
\en
\sk
We now use the twist to  
deform the commutative geometry on spacetime (vectorfields, 1-forms, 
exterior algebra, tensor algebra, symmetry algebras, covariant derivatives etc.) 
into the twisted noncommutative one. 
The guiding principle is the observation that every time we have 
a bilinear map $$\mu\,: X\times Y\rightarrow Z~~~~~~~~~~~~~~~~~~$$
where $X,Y,Z$ are vectorspaces, 
and where there is an action of the Lie algebra $g$   
(and therefore of $\FF^{-1}$) on $X$ and $Y$
we can combine this map with the action of the twist. In this way
we obtain a deformed version $\mu_\st$ of the initial bilinear map $\mu$:
\eqa
\mu_\st:=\mu\circ \FF^{-1}~,\label{generalpres}&~~~~~~~~~~~~~~&
\ena
{\vskip -.8cm}
\eqa
{}~~~~~~~~~~~~~\mu_\st\,:X\times  Y&\rightarrow& Z\nn\\
(\x, \y)\,\, &\mapsto& \mu_\st(\x,\y)=\mu(\of^\al(\x),\of_\al(\y))\nn~.
\ena
The cocycle condition (\ref{propF1})
implies that if $\mu$ is an associative product then also $\mu_\st$ is an associative product.
\sk\noi
{\bf Algebra of Functions $A_\st$}. If $X=Y=Z=Fun(M)$ where $A\equiv Fun(M)$ is the space of functions on spacetime 
$M$,
we obtain the star-product formulae (\ref{starprodf}), (\ref{fhfg}).
The $\st$-product is associative because of the cocycle condition (\ref{propF1}).
We denote by $A_\st$ the noncommutative algebra of functions with the $\st$-product.
Notice that to define the $\st$-product we need condition II), the action of 
the Lie algebra on functions. In this case it is obvious; the 
action of $\partial_\mu$ on a function $h$ is just $\partial_\mu h$, i.e. the 
Lie derivative of $\partial_\mu$ on $h$. In the sequel we will always use the 
Lie derivative action.

\sk
\noi {\bf Vectorfields $\Xi_\st$}. We now deform the product 
$\mu : A\otimes \Xi\rightarrow \Xi$ between 
the space $A=Fun(M)$ of functions on spacetime $M$ and vectorfields. 
A generic vectorfield is 
$v=v^\nu\partial_\nu$. Partial derivatives acts on vectorfields via the 
Lie derivative action
\eq\label{onlyconst}
\partial_\mu(v)=[\partial_\mu,v]=\partial_\mu(v^\nu)\partial_\nu~.
\en
According to (\ref{generalpres}) the product 
$\mu : A\otimes \Xi\rightarrow \Xi$
is deformed into the product 
\eq
h\st v=\of^\al(h) \of_\al(v)~.
\en
Iterated use of (\ref{onlyconst}) gives
\eq
h\st v=\of^\al(h) \of_\al(v)
=\of^\al(h)\of_\al(v^\nu) \partial_\nu=
(h\st v^\nu)\partial_\nu ~.
\en
It is then easy to see that  $h\st (g\st v)=(h\st g)\st v$. We have thus 
constructed the $A_\st$ module of vectorfields. We denote it by $\Xi_\st$. 
As vectorspaces $\Xi=\Xi_\st$, but $\Xi$ is an $A$ module while $\Xi_\st$ is 
an $A_\st$ module.
\sk
\noi {\bf 1-forms $\Om_\st$}. 
The space of 1-forms $\Om$ becomes also an $A_\st$ module,
with the product between functions and 1-forms given again by following 
the general prescription (\ref{generalpres}): 
\eq
h\st\om :=\of^\al(h) \of_\al(\om)~.
\en
The action of $\of_\al$ on forms is given by iterating the Lie derivative 
action of the vectorfield $\partial_\mu$ on forms. Explicitly, if 
$\om=\om_\nu dx^\nu$ we have
\eq
\partial_\mu(\om)=\partial_\mu(\om_\nu)dx^\nu
\en
and $\om=\om_\nu dx^\nu=\om_\mu\st dx^\mu$.

Functions can be multiplied from the left or from the right,
if we deform the multiplication from the right we obtain the new product  
\eq
\om\st h := \of^\al(\om)\of_\al(h)
\en
and we ``move $h$ to the right" with the help of the $R$-matrix,
\eq
\omega\st h={\oR^\al}(h)\st \oR_{\al}(\om)~.
\en
We have defined the $\AAs$-bimodule of 1-forms.

\sk

\noi {\bf Tensorfields {$\TT_\st$}}.  Tensorfields form an algebra with 
the tensorproduct $\otimes$. We define $\TT_\st$ to be the noncommutative 
algebra of tensorfields. As vectorspaces $\TT=\TT_\st$ the noncommutative 
tensorproduct is obtained by applying (\ref{generalpres}):
\eq\label{defofthetensprodst}
\tau\otimes_\st\tau':=\of^\al(\tau)\otimes \of_\al(\tau')~.
\en 
Associativity of this product follows from the cocycle condition 
(\ref{propF1}).

\sk

\noi {\bf Exterior forms 
$\Omega^{\mbox{\boldmath $\cdot$}}_\st=\oplus_p\Omega^{p}_\st$}.
Exterior forms form an algebra wth product 
$\wedge :\,\Omega^{\mbox{\boldmath $\cdot$}}\times 
\Omega^{\mbox{\boldmath $\cdot$}}\rightarrow \Omega^{\mbox{\boldmath $\cdot$}}$.
We $\st$-deform the wedge product into the $\st$-wedge product,
\eq\label{formsfromthm}
\vartheta\wedge_\st\vartheta':=\of^\al(\vartheta)\wedge \of_\al(\vartheta')~.
\en 
We denote by $\Omega^{\mbox{\boldmath $\cdot$}}_\st$ 
the linear space of forms equipped with the wedge product  
$\wedge_\st$.

As in the commutative case exterior forms are totally 
$\st$-antisymmetric contravariant tensorfields. 
For example the 2-form 
$\omega\wedge_\st\omega'$ is the $\st$-antisymmetric combination
\eq\label{stantisymm}
\omega\wedge_\st\omega'= \omega\otimes_\st\omega'
-\oR^\al(\omega')\otimes_\st \oR_{\al}(\omega)~.
\en

The usual exterior derivative 
$d:A\rightarrow \Om$ satisfies the Leibniz rule $d(h\st g)=dh\st g+h\st
dg $ and is therefore also the $\st $-exterior derivative. 

\sk
\noi{\bf $\st$-Pairing between 1-forms and vectorfields}. 
We now consider the bilinear map 
\eqa\label{lerestcomm}
\le~,~\re : \,
\Xi\times \Omega_\st &\rightarrow & A~,\\
(v,\omega)~&\mapsto &\le v,\omega\re= 
\le v^\mu\partial_\mu,\omega_\nu dx^\nu\re=v^\mu\om_\mu~.
\ena
Always according to the general prescription (\ref{generalpres}) we deform 
this pairing into
\eqa\label{lerest}
\le~,~\re_\st : \,
\Xi_\st\times \Omega_\st &\rightarrow & A_\st~,\\
(\xi,\omega)~&\mapsto &\le \xi,\omega\re_\st
:=\le\of^\al(\xi),\of_\al(\omega)\re~.
\ena
It is easy to see that the $\st$-pairing satisfies the  
$A_\st$-linearity properties  
\eq
\le h\st u,\omega\st k\re_\st=h\st\le u,\omega\re_\st\st k~,
\en
\eq\label{linearp}
\le u, h\st\omega \re_\st=
{\oR^\al}(h)\st\le {\oR_{\al}}(u),\omega \re_\st~.
\en
Notice that $\le \partial_\mu,dx^\nu\re_\st=\le 
\partial_\mu,dx^\nu\re=\delta_\mu^\nu$.

Using the pairing $\le~\,,~\,\res$ we associate to any $1$-form
$\om$ the left $A_\st$-linear map $\le~\,,\om\res$. 
Also the converse holds: any left $A_\st$-linear map 
$\Phi:\Xis\rightarrow \AAs$ is of the form $\le~\,,\om\res$
for some $\omega$ (explicitly $\omega=\Phi(\partial_\mu) dx^\mu$). 
\sk
\noi {\bf $\st$-Hopf algebra of diffeomorphisms $U\Xi_\st$}. 
We recall that the (infinite dimensional) linear space $\Xi$ of smooth vectorfields on spacetime $M$ becomes a Lie algebra through the map
\begin{eqnarray}
[\quad ]: \quad\quad \Xi\times\Xi &\to& \Xi \nonumber\\
(u,v) &\mapsto& [u~v] .\label{2.1}
\end{eqnarray}
The element $[u~v]$ of $\Xi$ is defined by the usual Lie bracket
\begin{equation}
[u~v](h) = u(v(h)) - v(u(h)) ,\label{2.2}
\end{equation}
where $h$ is a function on spacetime.


The Lie algebra of vectorfields (i.e. the algebra of infinitesimal 
diffeomorphisms) can also be seen as an abstract Lie algebra without 
referring to the action of vectorfields on functions. 
The universal 
enveloping algebra $\UU$ 
of this abstract 
Lie algebra is the associative algebra (over $\mathbb{C}$) generated by the 
elements of $\Xi$ 
and the unit element $1$ and 
where the element $[u~v]$ is given by the commutator
$uv-vu$, i.e. $uv-vu=[u~v]$. Here $uv$ and $vu$ denotes the product in $\UU$.
The algebra $\UU$ is the universal enveloping algebra of infinitesimal diffeomorphisms, we shall denote its elements by the letter $\xi$, 
$\zeta$, $\eta$,\dots. 

The undeformed algebra $\UU$ has a natural Hopf algebra structure \cite{Sweedler}.
On the generators $u\in \Xi$ the coproduct map $\Delta$ the 
antipode and the counit $\epsi$ are defined by
\eqa\label{cosclass}
&&\Delta (u)=u\otimes 1 + 1\otimes u~,\nonumber\\
&&\varepsi(u)=0~,\\
&&S(u)=-u\nonumber~.
\ena
(and $\Delta(1)=1\otimes 1~,\varepsi(1)=1~,
S(1)=1$). The maps $\Delta$ and $\varepsilon$ are then extended
as algebra homomorphisms
and $S$ as antialgebra homomorphism to the full enveloping algebra,
$\Delta: \UU\rightarrow \UU\otimes \UU$, 
$\varepsi:\UU\rightarrow \mathbb{C}$ 
and $S:\UU\rightarrow \UU$, 
\begin{eqnarray}\label{cosmult}
\Delta (\xi\zeta) &:=& \Delta(\xi)\Delta(\zeta) ~,\nonumber\\
\varepsilon (\xi\zeta) &:=& \varepsilon(\xi)\varepsilon(\zeta)~ ,\label{2.5}\\
S(\xi\zeta) &:=& S(\zeta)S(\xi) ~.\nn
\end{eqnarray}
The extensions of $\D$, $\epsi$ and $S$ are well defined because they are compatible with the relations $uv-vu=[u~v]$ 
(for ex. $S(uv-vu)=S(v)S(u)-S(u)S(v)=-[u~v]=S[u~v]$).

On the generators, the coproduct encodes the Leibniz rule property $u(hg)=u(h)g+hu(g)$, the 
antipode expresses the fact that the inverse of the group element $e^u$
is $e^{-u}$, while the counit associates to every element $e^u$ the identity $1$.


\sk
In order to construct the deformed algebra of diffeomorphisms  
we apply the recepy (\ref{generalpres}) and deform the product in $\UU$ 
into the new product
\eq
\xi\st\zeta=\of^\al(\xi)\of_\al(\zeta)~.
\en
We call $\UU_\st$ the new algebra with product $\st$. 
As vectorspaces $\UU=\UU_\st$.
Since any sum of
products of vectorfields in $\UU$ can be rewritten as sum of
$\st$-products of vectorfields via the formula $u\,v=\f^\al(u)\st\f_\al(v)$, 
vectorfields $u$ generate the algebra $\UU_\st$.

It turns out \cite{GR2} that $\UU_\st$ has
also a natural Hopf algebra structure.
We describe it by giving the
coproduct, the counit and the antipode\footnote{Notice that because of the 
antisymmetry of $\theta^{\mu\nu}$ we have
$\oR^\al\!(u)\,\oR_\al=\oR_\al\,\,\; \oR^\al\!\!(u)$. Since 
$S_\st(\partial_\nu)=-\partial_\nu$ it is then easy to prove that 
$S_\st^2=id$. It is also easy to check that $\mu(S\otimes id)\D(u) = \mu(id \otimes S)\D(u) = \epsi(u)1=0$. This last property uniquely defines the antipode.} 
on the generators $u$ of $\UU_\st$:
\eqa\label{coproductu}
&&\D_\st (u)=u\otimes 1+ {\oR^\al}\otimes {\oR_{\al}}(u)~,\\[.3em]
&&\epsi_\st(u)=\epsi(u)=0~,\label{epsist}\label{counitpos}\\[.3em]
&&S_\st(u)=-\oR^\al(u)\,\oR_\al~.\label{antipostu}
\ena
In the appendix we prove for example coassociativity of the coproduct $\D_\st$.
We here show that the coproduct definition (\ref{coproductu}) can be inferred 
from a deformed Leibniz rule.

\sk
There is a natural action (Lie derivative) 
of $\Xi_\st$ on the space of functions $A_\st$.
It is given once again by combining the usual Lie derivative on functions 
$\ll_u(h)=u(h)$ with the twist $\FF$ as in (\ref{generalpres}),
\eq\label{stliederact}
\ll^\st_u(h):=\of^\al(u)(\of_\al(h))~.
\en
By recalling that every vectorfield can be written as 
$u=u^\mu\st\partial_\mu=u^\mu\partial_\mu$ we have
\eqa\nn
\ll^\st_u(h)&=&\of^\al(u^\mu\partial_\mu)(\of_\al(h))=\of^\al(u^\mu)\,\partial_\mu(\of_\al(h))\\[.3em]
&=&
u^\mu\st\partial_\mu(h)~,\label{exlu}
\ena
where in the second equality we have considered the explicit expression  (\ref{faexp}) of $\of^\al$ in terms of partial derivatives, and we have iteratively used the property 
$[\partial_\nu,u^\mu\partial_\mu]=\partial_\nu(u)\,\partial_\mu$. In the last equality we have used that the partial derivatives contained in $\of_\al$ commute with the partial derivative $\partial_\mu$.

In accordance with the coproduct formula (\ref{coproductu}) the differential operator $\ll^\st_u$ satisfies the deformed  Leibniz rule 
\eq
\ll^\st_u(h\st g)=\ll_u^\st(h)\st g + 
\oR^\al(h)\st \ll_{\oR_\al(u)}^\st(g)~.
\en
Indeed recalling that $u=u^\mu\st\partial_\mu=u^\mu\partial_\mu$ we have
\eqa
\ll^\st_u(h\st g)=
u^\mu\st \partial_\mu(h\st g)&=&u^\mu\st \partial_\mu(h)\st g 
+ u^\mu\st h\st\partial_\mu(g)\nn\\[.3em]&=&
\ll^\st_u(h)\st g 
+ \oR^\al(h)\st \oR_\al(u^\mu)\st\partial_\mu(g)\nn\\[.3em]
&=&\ll_u^\st(h)\st g 
+ \oR^\al(h)\st \ll_{\oR_\al(u)}^\st(g)~.
\ena

From (\ref{exlu}) it is also immediate to check the compatibility condition
\eq
\ll_{f\st u}^\st(h)=f\st\ll_u^\st(h)~,
\en
that shows that the action $\ll^\st$ is the one compatible with the
$A_\st$ module structure of vectorfields.
\sk
The action $\ll^\st$ of $\Xi_\st$ on $A_\st$ can be extended to all $\UU_\st$.
We recall that the action of $\UU$ on the space of functions 
can be defined by extending the Lie 
derivative. For any function $h\in A=Fun(M)$, we define the Lie derivative of a product of generators 
$u...vz$ in $\UU$ to be the compositon of the  Lie derivatives of the 
generators,
\eq
(u...vz)(h)=u(.\;.\;.\;v(z(h))...\!)~.
\en
Then by linearity we know the Lie derivative along any element $\xi$ of $\UU$. 
We then define
\eq
\ll^\st_\xi(h):={\of^\al(\xi)}(\of_\al(h))~.
\en
The map $\ll^\st$ is an action 
of $\UU_\st$ on $A_\st$, i.e. it represents the algebra $\UU_\st$ 
as differential operators on functions because
\eq
\ll^\st_u(\ll^\st_v(h))=\ll^\st_{u\st v}(h)~.
\en
\sk
\noi {\bf $\st$-Lie algebra of vectorfields $\Xi_\st$}. 
We now turn our attention to the issue of determining the Lie algebra $\Xi_\st$ of $\UU_\st$.
In the undeformed case the Lie algebra of the universal enveloping algebra 
$\UU$ is the linear 
subspace $\Xi$ of $\UU$ of primitive elements, i.e. of elements $u$ that have 
coproduct: 
\eq
\Delta(u)=u\otimes 1 +1\otimes u\ .
\en
Of course $\Xi$ generates $\UU$ and $\Xi$ is closed under the usual 
commutator bracket $[~,~]$,
\eq
[u,v]= u u-v u\in \Xi ~~~~\mbox{for all } u, v\in \Xi~.
\en
The geometric meaning of the bracket  $[u,v]$ is that it is the 
adjoint action of $\Xi$ on $\Xi$,
\eq\label{adactcomm}
[u,v]=ad_u\,v
\en
\eq\label{edfr}
ad_u\,v:=u_1vS(u_2)
\en
where we have used the notation $\D(u)=u_1\otimes u_2$, 
where a sum over $u_1$ and $u_2$ is understood. Recalling that 
$\D(u)=u\otimes 1+1\otimes u$ and that $S(u)=-u$, from (\ref{edfr})
we immediately obtain
(\ref{adactcomm}). In other words, the commutator $[u,v]$ 
is the Lie derivative of the left invariant vectorfield 
$u$ on the left invariant vectorfield $v$. 
More in general the adjoint action of $\UU$ on 
$\UU$ is given by 
\eq
ad_\xi\,\zeta=\xi_1\zeta S(\xi_2)~,
\en
where we used the notation (sum understood)
$$\D(\xi)=\xi_1\otimes\xi_2~.$$ 
For example $ad_{uv}\,\zeta=[u,[v,\zeta]]$. 
\sk
In the deformed case the coproduct is no more cocommutative and  
we cannot identify the Lie algebra of $\UU_\st$ with the primitive elements
of $\UU$, they are too few\footnote{This 
can already be seen at the semiclassical level, where we are left with the
symplectic structure. Primitive elements then correspond to
symplectic infinitesimal transformations.
Instead of restricting the set of transformations to those compatible with
the bivector  $\theta^{\mu\nu}$ we want to properly generalize/relax 
the notion of infinitesimal automorphism. In this way 
we do not consider $\theta^{\mu\nu}$ as the components of a bivector, but 
as a set of constant coefficients.
}. 
There are three natural conditions that according to
\cite{Woronowicz}
the $\st$-Lie algebra of $\UU_\st$
has to satisfy, see \cite{AC,SWZ}, and \cite{{AschieriTesi}} p. 41. 
It has to be a linear subspace $\Xi_\st$ of $\UU_\st$ such that
\eqa
i)&&\Xi_\st \mbox{ generates } \UU_\st~, \\
ii)&&\D_\st(\Xi_\st)\subset \Xi_\st\otimes 1+\UU_\st\otimes \Xi_\st ~,\\
iii)&&[\Xi_\st,\Xi_\st]_{\st}\subset \Xi_\st~.
\ena
Property $ii)$ implies a minimal deformation of the Leibnitz rule. 
Property $iii)$ is the closure of $\Xi_\st$ under the adjoint action:
\eq
[u,v]_{\st}=ad^\st_u\,v=u_{1_\st}\st v\st S(u_{2_\st})~,
\en
here we have used the coproduct notation $\D_\st(u)=u_{1_\st}\otimes u_{2_\st}$. 
More in general the adjoint action is given by
\eq
ad^\st_\xi\,\zeta:=\xi_{1_\st}\st \zeta \st S_\st(\xi_{2_\st})~,
\en
where we used the coproduct notation 
$\D_\st(\xi)=\xi_{1_\st}\otimes\xi_{2_\st}\,$.
\sk
In the case the deformation is given by a twist we have a natural 
candidate for the Lie algebra of the Hopf algebra $\UU_\st$.
We apply the recepy (\ref{generalpres}) and deform the Lie algebra product 
$[~\,~]$ given in (\ref{2.1}) into  
\begin{eqnarray}
[\quad ]_\st: \quad\quad \Xi\times\Xi &\to& \Xi \nonumber\\
(u,v) &\mapsto& [u~v]_\st:=[\of^\al(u)~\of_\al(v)]~ .\label{2.1st}
\end{eqnarray}
In $\UU_\st$ this $\st$-Lie bracket can be realized as a deformed commutator
\eqa 
[u~v]_\st&=&[\of^\al(u)~\of_\al(v)]=\of^\al(u)\of_\al(v)-\of_\al(v)\of^\al(u)
\nn\\
 &=&u\st v-\oR^\al(v)\st\oR_\al(u)~.
\ena
It is easy to see that the bracket  $[~\,~]_\st $ 
has the $\st$-antisymmetry property
\eq\label{sigmaantysymme}
[u~v]_\st =-[\oR^\al(v)~ \oR_\al(u)]_\st~ .
\en
This can be shown as follows
$$[u~v]_\st =[\of^\al(u)~\of_\al(v)]=-[\of_\al(v)
~\of^\al(u)]=
-[\oR^\al(v)~ \oR_\al(u)]_\st~. $$
A $\st$-Jacoby identity can be proven as well
\eq
[u ~[v~z]_\st ]_\st =[[u~v]_\st ~z]_\st  
+ [\oR^\al(v)~ [\oR_\al(u)~z]_\st ]_\st ~.
\en
The appearence of the $R$-matrix $\RR^{-1}=\oR^\al\otimes\oR_\al$ is not
unexpected. We have seen that $\RR^{-1}$ encodes the noncommutativity 
of the $\st$-product $h\st g=\oR^\al(g)\st\oR_\al(h)$ 
so that $h\st g$ do $\RR^{-1}\!$-commute. Then it is natural to define  
$\st$-commutators using the $\RR^{-1}$-matrix. In other words, 
the representation of the permutation group to be used 
on twisted noncommutative spaces is the one given by the $\RR^{-1}$ matrix.
\sk
We now show that the  subspace $\Xi_\st$ (that as vectorspace equals $\Xi$) has all 
the three properties $i),~ii),~iii)$. It satisfies $i)$ because any sum of
products of vectorfields in $\UU$ can be rewritten as sum of
$\st$-products of vectorfields via the formula $u\,v=\f^\al(u)\st\f_\al(v)$, 
and therefore $\st$-vectorfields generate the algebra. It obviously
satisfies $ii)$, and finally in the appendix we prove that it
satisfies $iii)$ by showing that the bracket $[u~v]_\st$ is indeed the adjoint 
action, $ad^\st_uv=[u~v]_\st$. 
\sk
We stress that the geometrical --and therefore physical-- interpretation of 
$\Xi_\st$ as infinitesimal diffeomorphisms  is due to the deformed Leibniz rule
property $ii)$ and to the closure of $\Xi_\st$ under the adjoint action. 
Property $ii)$ will be fundamental in order to define covariant derivatives
(cf. (\ref{ddsDuhv})).

\sk
\noi {\sl Note 1. } The Hopf algebra $\UU_\st$ can be described via the generators 
$X_u:=\f^\al(u)\f_\al$ rather than via the $u$ generators. 
The action of $X_u$ on functions is the differential 
operator
$X^\st_u\equiv\ll^\st_{X_u}$, we have
$X^\st_u(f)\equiv\ll^\st_{X_u}(f)=u(f)$, compare with eq. (5.2) in
\cite{G1}, see also \cite{proc}.
The generators $X_u$ satisfy the commutation relations 
$X_u\st X_v-X_v\st X_u=X_{[u,v]}$ and their coproduct is 
$\D_\st(X_u)=\FF(X_u\otimes 1+1\otimes X_u)\FF^{-1}$. We see that 
$\UU_\st$ is the abstract Hopf algebra of diffeomorphisms considered in 
\cite{G1}, end of Section 5. Since the elements $X_u$ generate $\UU_\st$,
invariance under the diffeomorphisms algebra $\UU_\st$
is equivalently shown by proving invariance under the $X_u$ or the $u$
generators.
Since $X_{\partial_\mu}=\partial_\mu$ partial derivatives belong to both 
sets of generators. We also have 
$\ll^\st_{\partial_\mu}(f)=\partial_\mu (f)=\partial^\st_\mu\triangleright f$.

\subsection{Relation between $\UU_\st$ and $\UU^\FF$}
In the previous four pages, using the twist $\FF$ and the 
general prescription (\ref{generalpres}) we have described the Hopf algebra
$$
(\UU_\st,\st,\D_\st,S_\st,\epsi)
$$
and its Lie algebra $(\Xi_\st, [~~]_\st)$. These are
a deformation of the cocommutative Hopf algebra
$$(\UU,\cdot,\D,S,\epsi)$$
and its Lie algebra $(\Xi, [~~]\,)$.
Usually given a twist $\FF$ one deforms the Hopf algebra 
$(\UU,\cdot\,\D,S,\epsi)$ into the Hopf algebra
$$
(\UU^\FF,\cdot,\D^\FF,S^\FF,\epsi)
$$
where the coproduct is deformed via
\eq
\D^\FF(\xi):=\FF\D(\xi)\FF^{-1}~,
\en
while product, antipode and counit are undeformed 
$\cdot^\FF=\cdot~,S^\FF=S~,\epsi^\FF=\epsi\,$ 
($S^\FF=S$ only for abelian antisymmetric twists).
\sk
The Hopf algebras $\UU_\st$ and $\UU^\FF$ 
are isomorphic. As vectorspaces $\UU_\st=\UU=\UU^\FF$. The Hopf algebra 
isomorphism is given by the linear map $D:\UU^\st\rightarrow\UU^\st$ 
\eq
D(\xi)=\of^\al(\xi)\of_\al~.
\en
The inverse of the map $D$ is 
$$D^{-1}\equiv X:\, \xi\longmapsto X_\xi=\f^\al(\xi)\f_\al~,$$ 
indeed
$D(X_\xi)=\of^\be(\f^\al(\xi)\f_\al)\of_\be=
\of^\be(\f^\al(\xi))\f_\al\of_\be=(\of^\be\f^\al)(\xi)\,\f_\al\of_\be=\xi\,$
where we used that partial derivatives commute among themselves and 
in the last line we used $\FF^{-1}\FF=1\otimes 1$.
Explicitly the Hopf algebra isomorphisms between $\UU^\st$ and $\UU^\FF$ is, 
\cite{GR2}
\eqa \label{D alg-homo}
&&D(\xi\st \zeta)=D(\xi)D(\zeta)~,\\
&&\D_\st =(D^{-1}\otimes D^{-1})\circ \D^\FF\circ D~,\label{DST}\\
&&S_\st =D^{-1}\circ S^\FF \circ D~.
\ena
Under this isomorphism the Lie algebra $\Xi_\st$ is mapped into the Lie algebra
$\Xi^\FF:=D({\Xi_\st})$ of all elements $$u^\FF:=D(u)=\of^\al(u)\of_\al~.$$
The bracket in $\Xi^\FF$ is the deformed commutator
\eq
[u^\FF,v^\FF]_\FF=u^\FF v^\FF -\oR^\al(v^\FF)\oR_\al(u^\FF)
\en
and it equals the adjoint action in $\UU^\FF$,
\eq
[u^\FF,v^\FF]_\FF=ad^\FF_{u^\FF}v^\FF=u^\FF_{1_\FF} \,v\, S(u^\FF_{2_\FF})~,
\en
where we used the notation $\D^\FF(\xi)=\xi_{1_\FF}\otimes \xi_{2_\FF}~.$
The usual Lie algebra $\Xi$ of vectorfields with the usual bracket $[u,v]=uv-vu$
is not properly a Lie algebra of $\UU^\FF$ because the commutator fails to be the 
adjoint action and the Leibniz rule is not of the type $ii)$. 
In particular the vectofields $u$ have not the geometric 
interpretation of infinitesimal diffeomorphisms.
\sk

\sk

\subsection{$\st$-Diffeomorphisms Symmetry}
In the commutative case the diffeomorphisms algebra $\UU$ acts on the
algebra of functions and more in general on the algebra of tensorfields 
via the Lie derivative. The Rimemann curvature, the Ricci tensor and
the curvature scalar are tensors and therefore they transform
covariantly  under the diffeomorphisms action. In the twisted case,
the $\st$-diffeomorphisms algebra $\UU_\st$ acts on the $\st$-algebra of
functions $A_\st$ and  more in general on the $\st$-algebra of
tensorfields $\TT_\st$. The action on functions is given by the 
$\st$-Lie derivative
defined in (\ref{stliederact}). 
Similarly the action on tensors is
given, according to (\ref{generalpres}), by
\eq\label{stliederacten}
\ll^\st_u(\tau):=\of^\al(u)(\of_\al(\tau))~.
\en
This expression defines an action because 
$
\ll^\st_u(\ll^\st_v(\tau))=\ll^\st_{u\st v}(\tau)\,.$ 
In particular the $\st$-Lie derivative is a representation of the
Lie algebra of infinitesimal diffeomorphisms $\Xi_\st$,
\eq
\ll^\st_u\,\ll^\st_v-\ll^\st_{\oR^\al(v)}\, \ll^\st_{\oR_\al(u)}
=\ll^\st_{[u~v]_\st}~,
\en
where $\ll^\st_u\,\ll^\st_v=\ll^\st_u\circ\ll^\st_v$ i
s the usual composition of 
operators. The coproduct in $\UU$ is compatible with the product in
the tensorfields algebra because
\eq
\ll^\st_u(\tau\otimes_\st \tau')=\ll_u^\st(\tau)\st \tau' + 
\oR^\al(\tau)\st \ll_{\oR_\al(u)}^\st(\tau')~.
\en

In Section 4 we introduce the noncommutative Riemann tensor and  
Ricci curvature,
and  show that they are indeed tensors. Then they transform covariantly
under the action of the $\st$-diffeomorphism algebra.
The corresponding noncommutative Einstein equations satisfy the symmetry 
principle of noncommutative general covariance, 
i.e. they are covariant under $\st$-diffeomorphism symmetry.


\section{Poincar\'e Symmetry}
The considerations about the undeformed Hopf algebra $\UU$, and the
Hopf algebras $\UU_\st$ and $\UU^\FF$ hold independently from
$\Xi$ being the Lie algebra of infinitesimal diffeomorphisms. 
In this section we study the case of the deformed Poincar\'e algebra. 
It can be seen as an abstract algebra or also as a subalgebra of 
infinitesimal diffeomorphisms $\Xi$. 

\subsection{$\st$-Poincar\'e algebra}
We start by recalling that the usual Poincar\'e Lie algebra $iso(3,1)$:
\eqa\label{LiePoinc} [P_\mu,P_\nu]&=&0\ ,\nn\\[.3em]
[P_\rho , M_{\mu\nu}]&=&i(\eta_{\rho\mu}P_\nu-\eta_{\rho\nu}P_\mu)\ ,\\[.3em]
[M_{\mu\nu},M_{\rho\sigma}]&=&-i(\eta_{\mu\rho}
M_{\nu\sigma}-\eta_{\mu\sigma}M_{\nu\rho}
-\eta_{\nu\rho}M_{\mu\sigma}+\eta_{\nu\sigma}M_{\mu\rho})\ ,
\ena
is not a symmetry of $\theta$-noncommutative space because 
the relations 
\eq
x^\mu\st x^\nu -x^\nu\st x^\mu =i\theta^{\mu\nu}
\en
are not compatible with  Poincar\'e transformations. Indeed consider the
standard representation of the Poincar\' e algebra on functions $h(x)$,
\eq\label{representation}  {P}_\mu (h) =i\partial_\mu (h)\ ,\ {M}_{\mu\nu}
(h)=i(x_\mu\partial_\nu - x_\nu\partial_\mu)(h)\ ,\en
then we have
$M_{\rho\sigma}(\theta^{\mu\nu})=0$
while $M_{\rho\sigma}(x^\mu\st x^\nu -x^\nu\st x^\mu)\not=0$. 
This is so because we use the undeformed Leibniz rule 
$M_{\rho\sigma}(x^\mu\st x^\nu -x^\nu\st x^\mu)=
M_{\rho\sigma}(x^\mu)\st\, x^\nu+ x^\mu\st M_{\rho\sigma}(x^\nu)$. 
In other words the Hopf algebra $U(iso(3,1))$ generated by the Poincar\'e
Lie algebra and with usual coproducts
\eq
\D(P_{\mu})=P_{\mu}\otimes 1 +1\otimes P_{\mu}~~,~~~
\D(M_{\mu\nu})=M_{\mu\nu}\otimes 1 +1\otimes M_{\mu\nu}
\en
is not a symmetry of noncommutative spacetime. 
\sk
One approach to overcome this problem is to just deform the coproduct $\D$
into the new coproduct $\D^\FF(M_{\mu\nu})=\FF\D(M_{\mu\nu})\FF^{-1}$
(see  next subsection).

Another approach is to observe first that the action of $M_{\rho\sigma}$ on 
$h\st g$ is hybrid, indeed it mixes ordinary products 
with $\st$-products: 
$M_{\mu\nu}(h\st g)=ix_\mu\partial_\nu(h\st g)
-ix_\nu\partial_\mu(h\st g)$.  This is cured by
considering a different action of the generators 
$P_\mu$ and $M_{\mu\nu}$ on noncommutative spacetime. 
The $\ll^\st$  action  defined in (\ref{stliederact}), accordingly with the general 
prescription  (\ref{generalpres}),  exactly replaces 
the ordinary product with the $\st$-product. For any function 
$h(x)$ we have, 
\eqa
\ll^\st_{P_\mu}(h)~
&=&i\partial_\mu (h)\nn\\
\ll^\st_{M_{\mu\nu}}(h)
&=&ix_\mu\st\partial_\nu (h) - ix_\nu\st\partial_\mu (h)~.\label{actstP}
\ena
This action of the Poincar\'e generators on functions can be extended to
an action of the universal enveloping algebra $\UP$ if $\UP$ 
is endowed with the new $\st$-product 
\eqa\label{stprodPoi}
\xi\st\zeta&:=&\of^\al(\xi)\of_\al(\zeta)~\\[.3em]
\nn&=&\sum {1\over{n!}}\left( -i\over 2\right)^n\theta^{\rho_1\sigma_1}
\ldots\theta^{\rho_n\sigma_n}
[P_{\rho_1}\ldots [P_{\rho_n},\xi]...]\,
[P_{\sigma_1}\ldots [P_{\sigma_n},\zeta]...]\,,
\ena
for all $\xi$ and $\zeta$ in $\UP$. For example it is easy to see that
\eq
\ll^\st_{M_{\mu\nu}\st M_{\rho\sigma}}(h)=\ll^\st_{M_{\mu\nu}}
(\ll^\st_{M_{\rho\sigma}}(h))~.
\en
In formula (\ref{stprodPoi})
we have identified the Lie algebra of partial derivatives 
with the Lie algebra of momenta $P_\mu$, so that
\eq
\FF=e^{{i\over 2}\theta^{\mu\nu}{P_\mu}\otimes{P_\nu}}~~~,~~~
\RR^{-1}=e^{{i}\theta^{\mu\nu}{P_\mu}
\otimes{P_\nu}}~.
\en 
This identification is uniquely fixed by the representation 
(\ref{representation}): $P_\mu=i\partial_\mu$. 
Since products of the generators $P_\mu$ and $M_{\mu\nu}$ can be rewritten as sum of
$\st$-products  via the formula $\xi\zeta=\f^\al(\xi)\st\f_\al(\zeta)$,
the elements $P_\mu$ and $M_{\mu\nu}$ generate the algebra $U_\st(iso(3,1))$.
\sk
The coproduct compatible with noncommutative spacetime is inferred from the 
Leibniz rule 
\eqa
x_\mu\st\partial_\nu(h \st g)&=&x_\mu\st\partial_\nu(h)\st g 
+ x_\mu\st h\st\partial_\nu(g)\nn\\
&=&x_\mu\st\partial_\nu(h)\st g + \oR^\al(h)\st \oR_\al(x_\mu)\st\partial_\nu(g)~.
\ena
The coproduct that implements this Leibniz rule is (cf. (\ref{coproductu}))
\eq
\D_\st(M_{\mu\nu})=M_{\mu\nu}\otimes 1+\oR^\al\otimes \oR_\al(M_{\mu\nu})~.
\en
Explicitly the coproduct on the generators $P_\mu$ and $M_{\mu\nu}$ reads
\eqa
\Delta_\st(P_\mu)&=&P_\mu\otimes1+1\otimes
P_\mu~,\nn\\[.3em]
\label{copPoinst} 
\Delta_\st(M_{\mu\nu})&=&
M_{\mu\nu}\otimes 1+1 \otimes
M_{\mu\nu}+i\theta^{\alpha\beta}P_\al\otimes [P_\be,M_{\mu\nu}]~.
\ena
The counit and antipode on the generators can be calculated from  
(\ref{counitpos}) and (\ref{antipostu}), 
they are 
\eq
\epsi(P_\mu)=\epsi(M_{\mu\nu})=0~~,~~~
S_\st(P_\mu)=-P_\mu~~,~~~S_\st(M_{\mu\nu})=-M_{\mu\nu}
-i\theta^{\rho\sigma}[P_\rho,M_{\mu\nu}]P_\sigma~.
\en
We have constructed the Hopf algebra $U_\st(iso(3,1))$.
\sk

We recall that there are three natural conditions that the $\st$-Poincar\'e Lie 
algebra $iso_\st(3,1)$
has to satisfy. It has to be a linear subspace of $U_\st(iso(3,1))$ such that
if $\{t_i\}_{i=1,...n}$ is a basis of $iso_\st(3,1)$, we have (sum understood on repeated indices)
\eqa
i)&~&\{t_i\} \mbox{ generates } U_\st(iso(3,1)) \nn\\[.2em]
ii)&~&\D_\st(t_i)= t_i\otimes 1+f_i{}^j\otimes t_j \nn\\[.2em]
iii)&~&[t_i,t_j]_\st=C_{ij}{}^kt_k\nn
\ena
where $C_{ij}{}^k$ are structure constants and $f_i{}^j\in U^\FF(iso(3,1))$ 
($i,j=1,...n$).
%
In the last line the bracket $[~,~]_{\st}$ is the adjoint action
(we use the notation $\D_\st(t)=t_{1_\st}\otimes t_{2_\st}$)$ \,$:
\eq\label{adac123}
[t,t']_{\st}:=ad^\st_t\,t'=t_{1_\st}\st t'\st S_\st(t_{2_\st})~.
\en
We have seen that 
the elements $P_\mu$ and $M_{\mu\nu}$ generate $U_\st(iso(3,1))$.
They are deformed {\sl infinitesimal} generators because they satisfy 
the Leibniz rule $ii)$ and
because they close under the adjoint action $iii)$. In order to prove
property $iii)$ we perform a short calculation and obtain  
the explicit expression of the adjoint action (\ref{adac123}), 
\eqa
[P_\mu,P_\nu]_{\st}&=&[P_\mu,P_\nu]\,,\nn\\[.3em]
[P_\rho , M_{\mu\nu}]_{\st}&=&[P_\rho , M_{\mu\nu}]=-[M_{\mu\nu}, P_\rho]_{_\st}\,,\nn\\[.3em]
[M_{\mu\nu},M_{\rho\sigma}]_{\st}&=&
M_{\mu\nu}\st M_{\rho\sigma}-M_{\rho\sigma}\st M_{\mu\nu} 
-i\theta^{\al\be}[P_\al,M_{\rho\sigma}][P_\be,M_{\mu\nu}]=
[M_{\mu\nu},M_{\rho\sigma}]\nn
~.
\ena
Notice that this result  shows that the adjoint action (\ref{adac123})
equals the
deformed commutator $$t\st t'-\oR^\al(t')\st\oR_\al(t)~.$$ 
Property $iii)$, i.e. closure under the adjoint action, explicitly reads
\eqa\label{LiePoincst} [P_\mu,P_\nu]_{\st}&=&0\ ,\nn\\[.3em]
[P_\rho , M_{\mu\nu}]_{\st}&=&i(\eta_{\rho\mu}P_\nu-\eta_{\rho\nu}P_\mu)
\,,\nn\\[.3em]
[M_{\mu\nu},M_{\rho\sigma}]_{\st}&=&
-i(\eta_{\mu\rho}M_{\nu\sigma}-\eta_{\mu\sigma}M_{\nu\rho}
-\eta_{\nu\rho}M_{\mu\sigma}+\eta_{\nu\sigma}M_{\mu\rho})\ .
\ena
We notice that the structure constants are the same as in the undeformed case,
however the adjoint action $[M_{\mu\nu},M_{\rho\sigma}]_{\st}$ 
is not the commutator $M_{\mu\nu}\st M_{\rho\sigma}-M_{\rho\sigma}\st M_{\mu\nu}$
anymore, it is a deformed commutator quadratic in the 
generators and antisymmetric.

From (\ref{LiePoincst}) we immediately obtain the Jacoby identities:
\eq
[t \,,[t',t'']_{_\st} ]_{_\st} +[t' \,,[t'',t]_{_\st} ]_{_\st} 
+ [t'' \,,[t,t']_{_\st} ]_{_\st} =0~,
\en
for all $t,t',t''\in iso_\st(3,1)$.
\sk
It can be proven that the Hopf algebra $U_\st(iso(3,1))$ is the algebra 
freely generated by $P_\mu$ and $M_{\mu\nu}$ (we denote the product by $\st$)
modulo the relations $iii)$.
\sk
\noi {\sl Note 2.} In \cite{ACPoincare} we studied quantum Poincar\'e
groups  (in any dimension) obtained via abelian 
twists $\FF$ different from the one considered here. 
Their Lie algebra is 
described according to $i),~ ii),~iii)$ 
(see for ex. eq. (6.65),(7.36),(7.6),(7.7) in the first paper 
in \cite{ACPoincare}). Because of these three properties
the Lie algebra defines a differential calculus on the quantum Poincar\'e group 
manifold that respects the quantum Poincar\'e symmetry (i.e. that is bicovariant).

\subsection{Twisted Poincar\'e algebra}
The Poincar\'e Hopf algebra $U^\FF(iso(3,1))$ is another deformation of $\UP$.
As algebras 
$U^\FF(iso(3,1))=U(iso(3,1))$; but $U^\FF(iso(3,1))$ has the new coproduct	
\eq
\D^\FF(\xi)=\FF\D(\xi)\FF^{-1} ~,
\en
for all $\xi\in U(iso(3,1))$. 
In order to write the explicit expression for  $\D^\FF(P_\mu)$ and 
$\D^\FF(M_{\mu\nu})$, we use the Hadamard formula
$$Ad_{e^X}Y=e^X\ Y\
e^{-X}=\sum\limits_{n=0}^\infty\frac{1}{n!}\underbrace{[X,[X,...[}_n
X,Y]]=\sum\limits_{n=0}^\infty\frac{(ad X)^n}{n!}\ Y$$ and 
the relation $[P\otimes P', M\otimes 1]=[P,M]\otimes P'$, and thus obtain 
\cite{Chaichian}, \cite{Wess} 
\eqa
\Delta^\FF(P_\mu)&=&P_\mu\otimes1+1\otimes
P_\mu~,\nn\\[.3em]
\label{copPoin} 
\Delta^\FF(M_{\mu\nu})&=&M_{\mu\nu}\otimes 1+1 \otimes
M_{\mu\nu}\\& & ~-\frac{1}{2}\theta^{\alpha\beta}\left((\eta_{\alpha\mu}
P_\nu-\eta_{\alpha\nu}P_\mu)\otimes P_\beta
+P_\alpha\otimes(\eta_{\beta\mu}P_\nu-\eta_{\beta\nu}P_\mu)\right)
.\nn\ena
We have constructed the Hopf algebra $U^\FF(iso(3,1)$: it is the algebra 
generated by $M_{\mu\nu}$ and $P_\mu$ modulo the relations (\ref{LiePoinc}),
and with coproduct (\ref{copPoin}) and counit and 
antipode that are as in the undeformed case:
\eq
\epsi(P_\mu)=\epsi(M_{\mu\nu})=0~~,~~~
S(P_\mu)=-P_\mu~~,~~~S(M_{\mu\nu})=-M_{\mu\nu}~.
\en
This Hopf algebra is a symmetry of noncommutative spacetime provided that we
consider the ``hybrid'' action $M_{\mu\nu}(h\st g)=ix_\mu\partial_\nu(h\st g)
-ix_\nu\partial_\mu(h\st g)$. 

\sk
The Poincar\'e Lie algebra $iso^{\FF\!}(3,1)$
must be a linear subspace of $U^\FF(iso(3,1))$ such that
if $\{t_i\}_{i=1,...n}$ is a basis of $iso^{\FF\!}(3,1)$, we have (sum understood on repeated indices)
\eqa
i)&&\{t_i\} \mbox{ generates } U^\FF(iso(3,1)) \nn\\[.2em]
ii)&&\D^\FF(t_i)= t_i\otimes 1+f_i{}^j\otimes t_j \nn\\[.2em]
iii)&&[t_i,t_j]_\FF=C_{ij}{}^kt_k\nn
\ena
where $C_{ij}{}^k$ are structure constants and $f_i{}^j\in U^\FF(iso(3,1))$ 
($i,j=1,...n$).
%
In the last line the bracket $[~,~]_{_\FF}$ is the adjoint action:
\eq
[t,t']_{_\FF}:=ad^\FF_t\,t'=t_{1_\FF}t'S(t_{2_\FF})~.
\en
The statement that the {Lie algebra} of 
$U^\FF(iso(3,1))$ is the undeformed Poincar\'e Lie algebra (\ref{LiePoinc})
is not correct because conditions $ii)$
 and $iii)$ are not met by the generators $P_\mu$ and $M_{\mu\nu}$.
There is a canonical procedure in order to obtain the Lie algebra 
$iso^{\FF\!}(3,1)$ of 
$U^\FF(iso(3,1))$. We use 
the Hopf algebra isomorphism 
\eqa
D\,:\, U_\st(iso(3,1))&\rightarrow & U^\FF(iso(3,1))\nn\\[.3em]
            \xi&\mapsto& \of^\al(\xi)\of_\al\nn
\ena
and define $$iso^\FF(3,1):=D(iso_\st(3,1))~.$$
Explicitly consider the elements 
\eqa
P_\mu^\FF&:=&\of^\al(P_\mu)\of_\al=P_\mu~,\\[.3em]
M_{\mu\nu}^\FF&:=&\of^\al(M_{\mu\nu})\of_\al 
=M_{\mu\nu}-{i\over 2}
\theta^{\rho\sigma}[P_\rho,M_{\mu\nu}]P_\sigma
\nn\\&~&~~~~~~~~~~~~~~~=M_{\mu\nu}+{1\over 2}
\theta^{\rho\sigma}(\eta_{\mu\rho}P_\nu-
\eta_{\nu\rho}P_{\mu})P_\sigma
\ena
Their coproduct is 
\eqa
\Delta^\FF(P_\mu)&=&P_\mu\otimes1+1\otimes
P_\mu~,\nn\\[.3em]
\label{copPoinF} 
\Delta^\FF(M^\FF_{\mu\nu})&=&
M^\FF_{\mu\nu}\otimes 1+1 \otimes
M^\FF_{\mu\nu}+i\theta^{\alpha\beta}P_\al\otimes [P_\be,M_{\mu\nu}]~.
\ena
The counit and antipode are
\eq
\epsi(P_\mu)=\epsi(M^\FF_{\mu\nu})=0~~,~~~
S(P_\mu)=-P_\mu~~,~~~S(M^\FF_{\mu\nu})=-M^\FF_{\mu\nu}
-i\theta^{\rho\sigma}[P_\rho,M_{\mu\nu}]P_\sigma~.
\en
The elements $P^\FF_\mu$ and $M^\FF_{\mu\nu}$ are generators because they 
satisfy condition $i)$ (indeed $M_{\mu\nu}=M^\FF_{\mu\nu}+{i\over 2}
\theta^{\rho\sigma}[P_\rho,M^\FF_{\mu\nu}]P_\sigma$). They are deformed 
{\sl infinitesimal} generators because they satisfy the Leibniz rule $ii)$ 
and because they close under the Lie bracket $iii)$. Explicitly 
\eqa\label{LiePoincF} [P_\mu,P_\nu]_{_\FF}&=&0\ ,\nn\\[.3em]
[P_\rho , M^\FF_{\mu\nu}]_{_\FF}&=&i(\eta_{\rho\mu}P_\nu-\eta_{\rho\nu}P_\mu)
\,,\nn\\[.3em]
[M^\FF_{\mu\nu},M^\FF_{\rho\sigma}]_{_\FF}&=&
-i(\eta_{\mu\rho}M^\FF_{\nu\sigma}-\eta_{\mu\sigma}M^\FF_{\nu\rho}
-\eta_{\nu\rho}M^\FF_{\mu\sigma}+\eta_{\nu\sigma}M^\FF_{\mu\rho})\ .
\ena
We notice that the structure constants are the same as in the undeformed case,
however the adjoint action $[M^\FF_{\mu\nu},M^\FF_{\rho\sigma}]_{_\FF}$ 
is not the commutator anymore, it is a deformed commutator quadratic in the 
generators and antisymmetric:
\eqa
[P_\mu,P_\nu]_{_\FF}&=&[P_\mu,P_\nu]\,,\nn\\[.3em]
[P_\rho , M^\FF_{\mu\nu}]_{_\FF}&=&[P_\rho , M^\FF_{\mu\nu}]\,,\nn\\[.3em]
[M^\FF_{\mu\nu},M^\FF_{\rho\sigma}]_{_\FF}&=&
[M^\FF_{\mu\nu},M^\FF_{\rho\sigma}]
-i\theta^{\al\be}[P_\al,M_{\rho\sigma}][P_\be,M_{\mu\nu}]~.
\ena
From (\ref{LiePoincF}) we immediately obtain the Jacoby identities:
\eq
[t \,,[t',t'']_{_\FF} ]_{_\FF} +[t' \,,[t'',t]_{_\FF} ]_{_\FF} 
+ [t'' \,,[t,t']_{_\FF} ]_{_\FF} =0~,
\en
for all $t,t',t''\in iso^{\FF\!}(3,1)$.
\sk

\section{Covariant Derivative, Torsion and Curvature}\label{covderivative}
By now we have acquired enough knowledge on $\st$-noncommutative 
differential geometry to develop the formalism of covariant
derivative, torsion, curvature 
and Ricci tensors just by following the usual classical formalism.

We define a $\st$-covariant derivative 
$\dd^\star_u$ along the vector field $u\in \Xi$
to be a linear map $\dds_u:\Xis\rightarrow\Xis$ such that
for all $u,v,z\in\Xi_\st,~ h\in A_\st$:
\eqa
&&\dd_{u+v}^{\star}z=\dd_{u}^{\star}z+\dd_{v}^{\star}z~,\\[.35cm]
&&\dd_{h\star u}^{\star}v=h\star\dd_{u}^{\star}v~,\label{ddal}\\[.35cm]
&&\dd_{u}^{\star}(h\star v)
\,=\,\mathcal{L}_u^{\star}(h)\star v+
\oR^\al(h)\st\dd^\st_{\oR_\al(u)}v\label{ddsDuhv}
\ena
Notice that in the last line we have used the coproduct 
formula (\ref{coproductu}),
$\D_\st (u)=u\otimes 1+ \oR ^\al\otimes {\oR_{\al}}(u)
$. Epression (\ref{ddsDuhv}) is well defined because
$\oR_\al(u)$ 
is again a vectorfield.

\sk
The (noncommutative) connection coefficients 
${\Gamma_{\mu\nu}}^\sigma$ are given by 
\eq
\dds_{\mu}\partial_\nu=
{\Gamma_{\mu\nu}}^\sigma\st\partial_\sigma=
{\Gamma_{\mu\nu}}^\sigma\,\partial_\sigma~,
\en
where  $\dds_\mu=\dds_{\partial_\mu}$.
They uniquely determine the connection, indeed let 
$z=z^\mu\st\partial_\mu$,
$u=u^\nu \st\partial_\nu$, then
\eqa
\dds_z u&=&z^\mu\st\dds_\mu (u^\nu\st\partial_\nu)\nn\\
&=&z^\mu\st\partial_\mu(u^\nu)\,\partial_\nu+
z^\mu\st u^\nu\st\dds_\mu\partial_\nu\nn\\
&=&z^\mu\st\partial_\mu(u^\nu)\,\partial_\nu+
z^\mu\st u^\nu\st \Gamma_{\mu\nu}{}^\sigma\,\partial_\sigma\,\,;
\ena
these equalities are equivalent to the connection properties 
(\ref{ddal}) and (\ref{ddsDuhv}).

The covariant derivative is extended to tensorfields using the deformed 
Leibniz rule
$$\dds_u(v\otimes_\st z)= \dds_{u}(v)\ots z + \oR^\al(v)\ots 
\dds_{\oR_\al(u)}z\,\,.
$$
Requiring compatibility of the covariant derivative with the contraction
operator
gives the covariant derivative on 1-forms, we have $\dds_z=z^\mu\st\dds_\mu\,$, and
\eq
\dds_{\mu} (\om_\rho  dx^\rho)
=\partial_\mu(\om_\rho)\,dx^\rho-
\Ga_{\mu\rho}{}^\nu\st \om_\nu\: dx^\rho ~.
\en
\sk
The torsion $\tr$ and the curvature $\rr$ associated to
a connection $\dd^\st$ are the linear maps  
$\tr:\Xis \times \Xis\rightarrow\Xis$, and 
$\rr^\star : \Xis\times \Xis\times\Xis\rightarrow\Xis$ defined by
\eqa
\tr(u,v)&:=&\dd_{u}^{\star}v-\dd_{\oR^{\alpha}(v)}^{\star}\oR_{\alpha}(u)
-[u~v]_{\star}~,\\[.2cm]
\rr(u,v,z)&:=&\dd_{u}^{\star}\dd_{v}^{\star}z-
\dd_{\oR^\al{(v)}}^{\star}\dd_{\oR_\al(u)}^{\star}z-\dds_{[u\,v]_\st} z~,
\ena
for all $u,v,z\in\Xis$.
{}From  the antisymmetry property of the bracket $[\,~]_\st$, 
see (\ref{sigmaantysymme}), it
easily follows that 
the torsion $\tr$ and the curvature $\rr$ have the following 
$\st$-antisymmetry property
\eqa\label{Tantysymm}
\tr(u,v)&=&-\tr(\oR^\al(v),\oR_\al(u))~,\nn\\[.3em]
\rr(u,v,z)&=&-\rr(\oR^\al(v),\oR_\al(u),z)~.\nn
\ena
The presence of the $R$-matrix in the definition of torsion and curvature insures 
that  $\tr$ and $\rr$  are left $A_\st$-linear maps \cite{GR2},\cite{NG}, i.e.
$$
\tr(f\star u,v)=f\star \tr(u,v)~
~~,~~~\tr(\partial_\mu,f\star v)= f\st\tr(\partial_\mu ,v)
$$
(for any $\partial_\mu$), and similarly for the curvature.
We have seen that any left $A_\st$-linear map $\Xi_\st\rightarrow
A_\st$ is identified with a tensor, precisely a 1-form 
(recall comments after (\ref{linearp})),
similarly the $A_\st$-linearity of $\tr$ and $\rr$ insures that we
have well defined the torsion tensor and the curvature  tensor.

One can also prove (twisted) first and second Bianchi identities
\cite{GR2},\cite{NG}.
\sk
The coefficients  $\tr_{\mu\nu}{}^\rho$  and $\rr_{\mu\nu\rho}{}^\sigma$
with respect to the partial derivatives basis $\{\partial_\mu\}$ are defined by
\eq
\tr(\partial_\mu,\partial_\nu)=\tr_{\mu\nu}{}^\rho\partial_\rho~~~,~~~~~~
\rr(\partial_\mu,\partial_\nu,\partial_\rho)=
\rr_{\mu\nu\rho}{}^\sigma\partial_\sigma
\en 
and they explicitly read
\eqa
\tr_{\mu\nu}{}^\rho&=&\Ga_{\mu\nu}{}^\rho-\Ga_{\nu\mu}{}^\rho~~,\nn\\[.3em]
\rr_{\mu\nu\rho}{}^\sigma&=&\partial_\mu\Ga_{\nu\rho}{}^\sigma-
\partial_\nu\Ga_{\mu\rho}{}^\sigma
+\Ga_{\nu\rho}{}^\beta\st\Ga_{\mu\beta}{}^\sigma-
\Ga_{\mu\rho}{}^\beta\st\Ga_{\nu\beta}{}^\sigma~.
\ena

As in the
commutative case the Ricci tensor is a contraction of the curvature
tensor, 
\eq
\ric_{\mu\nu}={\rr_{\rho\mu\nu}}^\rho.
\en
A definition of the Ricci tensor that is independent from the $
\{\partial_\mu\}$ basis is also possible.

\section{Metric and Einstein Equations}\label{riemgeo}
In order to define a $\st$-metric we need to define $\st$-symmetric 
elements in $\Oms\ots\Oms$. Recalling the
$\st$-antisymmetry of the wedge $\st$-product (\ref{stantisymm}) 
we see that  $\st$-symmetric elements are of the form
\eq
\omega\otimes_\st\omega'
+\oR^\al(\omega')\otimes_\st \oR_{\al}(\omega)~.
\en
In particular any symmetric tensor in
$\Om\otimes\Om\,$,
\eq
g=g_{\mu\nu}dx^\mu\otimes dx^\nu~,
\en
$g_{\mu\nu}=g_{\nu\mu}$, is also a $\st$-symmetric tensor in 
 $\Oms\ots\Oms$ because
\eq
g=g_{\mu\nu}dx^\mu\otimes dx^\nu=g_{\mu\nu}\st dx^\mu\otimes_\st dx^\nu~
\en
and the action of the $R$-matrix is the trivial one on $dx^\nu$. 
We denote by $g^{\st \mu\nu}$ the star inverse of $g_{\mu\nu}$,
\eq
g^{\st \mu\rho}\st g_{\rho\nu}
=g_{\nu\rho}\st g^{\st \rho\mu}=\delta_\nu^\mu~.
\en
The metric $g_{\mu\nu}$ can be expanded order by order in the
noncommutative parameter $\theta^{\rho\sigma}$.
Any commutative metric is also a noncommutative metric, indeed the 
$\st$-inverse metric can be constructed order by order in the noncommutativity
parameter. Contrary to \cite{MadoreM}, \cite{MadoreFC}, 
we see that in our approach
there are infinitely many metrics compatible with 
a given noncommutative differential geometry, noncommutativity 
does not single out a preferred metric.

\sk
A connection that is metric compatible is a connection that for any vectorfield $u$ satisfies,
$\dds_u g=0$, this is equivalent to the equation
\eq
\dds_\mu g_{\rho\sigma}-\Ga_{\mu\rho}{}^\nu \st g_{\nu\sigma} -\Ga_{\mu\sigma}{}^\nu \st g_{\rho\nu}=0 ~.
\en
Proceeding as in the commutative case we obtain that there is a unique torsion 
free metric compatible connection \cite{G1}. It is given by
\eq
\Ga_{\mu\nu}{}^\rho={1\over 2}(\partial_\mu g_{\nu\sigma}+
\partial_\nu g_{\sigma\mu}-\partial_\sigma g_{\mu\nu})\st g^{\st \sigma\rho}
\en
\sk

We now construct the curvature tensor and the Ricci tensor using this 
uniquely defined connection. Finally the noncommutative Einstein equations 
(in vacuum) are
\eq\label{Einstein}
\ric_{\mu\nu}=0
\en
where the dynamical field is the metric $g$.
\sk
\sk\sk
\noi{\bf{\large Acknowledgements}}
\sk
\noi We would like to 
thank the organizers and the participants of the workshop for the fruitful 
and very pleasant atmosphere. 
We would also like to acknowledge fruitful discussions with the
participants of the WNG06 workshop on Noncommutative Geometry,
Cornelius University,  Bratislava June-July 2006.

Support from the Marie Curie European Reintegration Grant
MERG-CT-2004-006374 is acknowledged.
Partial support from the European Community's Human Potential Program
under contract MRTN-CT-2004-005104 and from the
Italian MIUR under contract PRIN-2003023852 is acknowledged.

\sk
\appendix
\section{Differential Operators and Vectorfields}
We briefly describe the algebra of differential operators and show that it 
is not a Hopf algebra by relating it to the Hopf algebra of infinitesimal 
diffeomorphisms.
\sk
Differential operator on the space of functions $A=Fun(\r4)$ 
are elements of the form 
$f(x)^{\mu_1\ldots\mu_n}\partial_{\mu_1}\ldots\partial_{\mu_n}$. They form
an algebra, the only nontrivial commutation relations are between
functions and partial derivatives,
\eq\label{defcommdiff}
\partial_\mu f=\partial_\mu(f) + f\partial_\mu~,
\en
where both $\partial_\mu$ and $f$ act on functions
(the action of 
$f$ on the function $h$ is given by the product $fh$.). 
Differential operators of zeroth order are functions. Differential
operators of first order \Diff$\,^1$ are derivations of the algebra
$A$ of functions (i.e. they satify the Leibniz rule) they are
therefore vectorfields $\Xi$ (infinitesimal diffeomorphisms). 
The isomorphism between vectorfields and first order differential
operators is given by the Lie derivative
\eqa
\ll :\,\Xi &\rightarrow& \Diff\,^1\nn\\[.3em]
        v&\mapsto& \ll_v
\ena
where $$\ll_v(f)=v(f)~.$$ We use the notation $\ll_v$ in order to stress
that the abstract Lie algebra element $v\in\Xi$ is seen as a
differential operator.
The Lie derivative can be extended to a map from the universal
enveloping algebra of vectorfields $\UU$ to all differential operators
\eqa\label{morphismsalg}
\ll :\,\Xi &\rightarrow& \Diff\,\nn\\[.3em]
        uv...z&\mapsto& \ll_u\,\ll_v\,...\ll_z
\ena
Notice that on the left hand side the product $uv$ is in $\UU$ (recall
the paragraph after (\ref{2.2})),
while on the right hand side the product
$\ll_u\,\ll_v=\ll_u\circ\,\ll_v$ 
is the usual composition product of operators. 

The map $\ll$ is an algebra morphism between the algebras $\UU$ and $\Diff$.
It is not surjective because the image of
$\UU$ does not contain the full space of functions $A$ but only
the constant ones (the multiples of the unit of $\UU$).

In order to show that the map $\ll\,:\,\UU\rightarrow \Diff\,$ is not
injective we consider the vectorfields
\eqa
u=f\partial_\mu ~&~,~&~~v=\partial_\nu~,\nn\\[.3em]
u'=f\partial_\nu
~&~,~&~~v'=\partial_\mu~,\nn
\ena
where for example $f=x^\nu$, and we show that 
\eq\label{diff9871}
~~~~~~~~~~~~~~~~~uv\not=u'v'~~~~~~~~~ \mbox{ in } \,\UU\,.   
\en
The map  $\ll$ is then not injective because 
$f\partial_\mu(\partial_\nu(h))=f\partial_\nu(\partial_\mu(h))$ for
any function $h$ implies
$$\ll_{uv}=
\ll_{u'v'}~.$$
The algebra $\UU$ is a Hopf algebra, in particular there is a well
defined coproduct map $\D$, and therefore one way to prove the inequalty
(\ref{diff9871}) is to prove that $\D(uv)\not=\D(u'v')$. We calculate
\eqa
\D(uv)&=&\D(u)\D(v)=uv\otimes 1+ u\otimes v +v\otimes u + 1\otimes uv\nn\\[.3em]
&=&f\partial_\mu\,\partial_\nu\otimes 1 +f\partial_\mu\otimes \partial_\nu
+\partial_\nu\otimes f\partial_\mu + 1\otimes f\partial_\mu\, \partial_\nu\nn~,
\ena
and 
\eq
\D(u'v')=f\partial_\nu\,\partial_\mu\otimes 1 +f\partial_\nu\otimes 
\partial_\mu
+\partial_\mu\otimes f\partial_\nu + 1\otimes f\partial_\nu\,
\partial_\mu\nn~.
\en
These two expressions are different. For example by applying the product map 
in $\UU$, $\cdot\,:\,\UU\otimes\UU\rightarrow \UU$  and then the map $\ll\,:\,\UU\rightarrow \Diff\,$ we respectively obtain
\eq
3 f\partial_\mu\,\partial_\nu+\partial_\nu\,f\partial_\mu \not=
3 f\partial_\nu\,\partial_\mu+\partial_\mu\,f\partial_\nu ~.\label{aqws}
\en 

{}From this proof we conclude that we cannot equip the algebra of differential operators 
\Diff$\;$ with a coproduct like the one in $\UU$.  The map defined by 
$\D(\ll_u)=\ll_u\otimes 1+1\otimes\ll_u$, and extended multiplicatively to all 
\Diff~ is not well defined because $\ll_{uv}=\ll_{u'v'}$, while 
$$\D(\ll_{uv})=\D(\ll_u)\D(\ll_v)\not=\D(\ll_{u'})\D(\ll_{v'})=\D(\ll_{u'v'})~,$$
as is easily seen by applying 
the product in \Diff$\:$, $\;\circ\,:\,\Diff\,\otimes\Diff\rightarrow\Diff~,$
(we obtain again (\ref{aqws})).
\section{Proof that the coproduct $\D_\st$ is coassociative}

We have to prove that 
$$(\D_\st\otimes id)\D_\st(u)=(id\otimes \D_\st)\D_\st(u)~.$$
The left hand side explicitly reads
\eqa
(\D_\st\otimes id)\D_\st(u)&=&(\D_\st\otimes id)(u\otimes 1+\oR^\al\otimes\oR_\al(u))\nn\\[.2em]&=&u\otimes 1\otimes 1 + \oR^\be\otimes\oR_\be(u)\otimes 1 +\D_\st(\oR^\al)\otimes \oR_\al(u)~.\nn
\ena
The right hand side is 
\eqa
(id\otimes\D_\st)\D_\st(u)&=&u\otimes \D_\st(1)+ 
\oR^\al\otimes\D_\st(\oR_\al(u))\nn\\[.2em]
&=& u\otimes 1\otimes 1+ \oR^\al\otimes\oR_\al(u)\otimes 1 + \oR^\al\otimes\oR^\ga\otimes \oR_\ga\oR_\al(u)~.\nn
\ena
These two expressions coincide because
\eqa\label{Dofar}
\D_\st(\oR^\al)\otimes \oR_\al&=&
e^{-i\theta^{\mu\nu}\D_\st(\partial_\mu)\otimes\partial_\nu}=
e^{-i\theta^{\mu\nu}(\partial_\mu\otimes 1\otimes\partial_\nu +
1\otimes\partial_\mu\otimes \partial_\nu)}\nn\\
&=&
\oR^\al\otimes\oR^\ga\otimes\oR_\ga\oR_\al~.
\ena
\section{Proof that the bracket $[u~v]_\st$ is the adjoint 
action}
We have to prove that 
$$[u~v]_\st=ad^\st_uv~.$$
We know that the backet  $[u~v]_\st$ equals the deformed commutator
$$
[u~v]_\st=u\st v -\oR^\al(v)\st\oR_\al(u)~.
$$
On the other hand, the adjoint action reads
\eq
ad^\st_uv=u_\1s\st v
\st S_\st(u_\2s)=u\st v + \oR^\al\st v\st S_\st(\oR_\al(u))
=u\st v - \oR^\al\st v\st \oR^\be(\oR_\al(u))\oR_\be\nn~.
\en
Now the property 
\eq\partial_\mu\st v=\partial_\mu \,v=\partial_\mu(v) 
+ v\,\partial_\mu\label{propparv}~,
\en 
that using the coproduct  
$\D_\st(\partial_\mu)\equiv\partial_{\mu_{\displaystyle{_\1s\!}}}\otimes
\partial_{\mu_{\displaystyle{_\2s}}}\!=\partial_\mu\otimes 1+
1\otimes\partial_\mu\,$ 
can be written as $$\partial_\mu\st v=\partial_\mu\, v=
\partial_{\mu_{\displaystyle{_\1s\!}}}(v)\:
\partial_{\mu_{\displaystyle{_\2s}}}~,$$ 
implies 
$$
\oR^\al\st v=\oR^\al v=\oR^\al_{\;\1s}(v)\,\oR^\al_{\;\2s}~.
$$
{}The coproduct formula (\ref{Dofar}) then implies
\eqa
\oR^\al\st v\st \oR^\be(\oR_\al(u))\oR_\be\nn&=& 
\oR^\al(v)\oR^\ga\st
\oR^\be ((\oR_\ga\oR_\al)(u))\oR_\be\nn\\[.3em]
&=&
\oR^\al(v)\st 
\oR^\be ((\oR_\ga\oR_\al)(u))\oR_\be\oR^\ga\nn\\[.3em]
&=&\oR^\al(v)\st 
(\oR^\be\oR_\ga\oR_\al)(u)\oR_\be\oR^\ga\nn\\[.3em]
&=&
\oR^\al(v)\st \oR_\al(u)\nn
\ena
where in the second equality we iterated property
(\ref{propparv}) (with $\oR^\be ((\oR_\ga\oR_\al)(u))\oR_\be$ insted of $v$)
and used the antysymmetry of $\theta^{\mu\nu}$ in order to cancel the
first addend in (\ref{propparv}). In the last equality we used that 
$\oR^\be\oR_\ga\otimes\oR_\be\oR^\ga=\RR^{-1}\RR=1\otimes 1$ because of the antysymmetry
of $\theta^{\mu\nu}$. 


\begin{thebibliography}{10}


\bibitem{Pauli:1985}
W.~Heisenberg,
\newblock {\em Letter from Heisenberg to Peierls},
\newblock in: W. Pauli, Scientific Correspondence, Vol. II, Berlin, Springer
  (1985).


\bibitem{Madore:1992}
J.~Madore,
\newblock {\em Gravity on fuzzy space-time},
\newblock Class. Quant. Grav. {\bf 9}, 69 (1992).



\bibitem{Castellani:1993ud}
  L.~Castellani,
 \newblock {\em  Differential calculus 
on $ISO_q(N)$, quantum Poincare algebra and $q$-gravity},
\newblock   Commun.\ Math.\ Phys.\  {\bf 171} (1995) 383
  hep-th/9312179,
%
 \newblock{\em The Lagrangian of q-Poincare gravity,}
\newblock  Phys.\ Lett.\ B {\bf 327} (1994) 22
  hep-th/9402033.



\bibitem{Doplicher1}
  S.~Doplicher, K.~Fredenhagen and J.~E.~Roberts,
  \newblock {\em The Quantum structure of space-time at the Planck scale and quantum
  fields},
\newblock   Commun.\ Math.\ Phys.\  {\bf 172} (1995) 187
  hep-th/0303037,
\newblock {\em  Space-time quantization induced by classical gravity,}
\newblock   Phys.\ Lett.\ B {\bf 331} (1994) 39.


\bibitem{CFF} A. Chamseddine, G. Felder, J. Fr\"ohlich,
{\sl Gravity in non-commutative geometry},
 Commun.Math.Phys. 155 (1993) 205 



\bibitem{ConnesG}
A.~Connes,
\newblock {\em Gravity coupled with matter and the foundation of non-
commutative geometry},
\newblock Commun. Math. Phys. {\bf 182}, 155 (1996), hep-th/9603053.


\bibitem{MadoreM}
J.~Madore and J.~Mourad,
\newblock {\em Quantum space-time and classical gravity},
\newblock J. Math. Phys. {\bf 39}, 423 (1998), gr-qc/9607060.


\bibitem{MajidJGP}
S.~Majid,
\newblock {\em Quantum and Braided group Riemannian geometry}
\newblock J. Geom. Phys., 30 113-146, 1999.


\bibitem{Moffat:2000gr}
J.~W. Moffat,
\newblock {\em Noncommutative quantum gravity},
\newblock Phys. Lett. {\bf B491}, 345 (2000), hep-th/0007181.


\bibitem{Chamseddine:2000si}
A.~H. Chamseddine,
\newblock {\em Deforming Einstein's gravity},
\newblock Phys. Lett. {\bf B504}, 33 (2001), hep-th/0009153.

\bibitem{Vacaru:2000yk}
S.~I. Vacaru,
\newblock {\em Gauge and Einstein gravity from non-Abelian gauge models on
 noncommutative spaces},
\newblock Phys. Lett. {\bf B498}, 74 (2001), hep-th/0009163.

\bibitem{Cardella:2002pb}
M.~A. Cardella and D.~Zanon,
\newblock {\em Noncommutative deformation of 
four dimensional 
Einstein  gravity},
\newblock Class. Quant. Grav. {\bf 20}, L95 (2003), hep-th/0212071.


\bibitem{G1}
  P.~Aschieri, C.~Blohmann, M.~Dimitrijevic, F.~Meyer, P.~Schupp and J.~Wess,
  {\em A gravity theory on noncommutative spaces,}
  Class.\ Quant.\ Grav.\  {\bf 22} (2005) 3511
  hep-th/0504183.

\bibitem{GR2}
  P.~Aschieri, M.~Dimitrijevic, F.~Meyer and J.~Wess,
  {\em Noncommutative geometry and gravity,}
  Class.\ Quant.\ Grav.\  {\bf 23} (2006) 1883,
  hep-th/0510059.

\bibitem{Szabo}
  R.~J.~Szabo,
  {\em Symmetry, gravity and noncommutativity,}
  hep-th/0606233.

\bibitem{Drinfeld1}   V.~G.~Drinfeld, 
{\em On constant quasiclassical solutions of the Yang-Baxter
  equations,}
\newblock Soviet Math. Dokl. {\bf 28} (1983) 667-671.

\bibitem{Drinfeld3}
  V.~G.~Drinfeld,
\newblock {\em   Quasi-Hopf Algebras,}
 \newblock  Lengingrad Math.\ J.\  {\bf 1} (1990) 1419
  [Alg.\ Anal.\  {\bf 1N6} (1989) 114].


\bibitem{Reshetikhin}
  N.~Reshetikhin,
  \newblock {\em Multiparameter Quantum Groups And Twisted Quasitriangular Hopf Algebras,}
\newblock   Lett.\ Math.\ Phys.\  {\bf 20} (1990) 331.



\bibitem{ACPoincare}
  P.~Aschieri and L.~Castellani,
  \newblock {\em Bicovariant Calculus on Twisted ISO(N), Quantum Poincare' Group and Quantum
  Minkowski Space,}
  Int.\ J.\ Mod.\ Phys.\ A {\bf 11} (1996) 4513
  q-alg/9601006,
  \newblock {\em R matrix formulation of the quantum inhomogeneous group ISO-q,r(N) and
  ISp-q,r(N),}
  Lett.\ Math.\ Phys.\  {\bf 36} (1996) 197
  hep-th/9411039.\\
  P.~Aschieri, L.~Castellani and A.~M.~Scarfone,
   {\em Quantum orthogonal planes: ISO (q,r) (n+1, n-1) and SO (q,r) (n+1, n-1)
  bicovariant calculi,}
  Eur.\ Phys.\ J.\ C {\bf 7} (1999) 159
  q-alg/9709032.





\bibitem{Kulish-Mudrov}
P.P.~Kulish, A.I.~Mudrov
  \newblock {\em  
 Twist-related geometries on q-Minkowski space}
Proc. Steklov Inst. Math. 226 (1999) 97-111, math.QA/9901019.

\bibitem{Lukierski}
  J.~Lukierski and M.~Woronowicz,
   {\em New Lie-algebraic and quadratic deformations of Minkowski space from
  twisted Poincare symmetries,}
  Phys.\ Lett.\ B {\bf 633} (2006) 116,
  hep-th/0508083.


\bibitem{Oeckl:2000eg}
R.~Oeckl,
\newblock {\em Untwisting noncommutative $R^d$ and the equivalence of quantum
  field theories},
\newblock Nucl. Phys. {\bf B581}, 559 (2000), hep-th/0003018.


\bibitem{Wess}
J.~Wess,
\newblock {\em Deformed coordinate spaces: Derivatives},
\newblock Lecture given at the Balkan workshop BW2003, August 2003, 
Vrnjacka Banja, Serbia , hep-th/0408080.

\bibitem{Chaichian}
M.~Chaichian, P.~Kulish, K.~Nishijima, and A.~Tureanu,
\newblock {\em On a Lorentz-invariant 
interpretation of noncommutative
  space-time and its implications on noncommutative QFT},
\newblock Phys. Lett. {\bf B604}, 98 (2004), hep-th/0408069.

\bibitem{Koch}
F.~Koch and E.~Tsouchnika,
\newblock {\em Construction of $\theta$-Poincare algebras and their invariants
  on $M_{\theta}$},
\newblock Nucl.\ Phys.\ B {\bf 717} (2005) 387
  hep-th/0409012.


\bibitem{Gonera:2005hg}
  C.~Gonera, P.~Kosinski, P.~Maslanka and S.~Giller,
  \newblock {\em Space-time symmetry of noncommutative field theory},
  Phys.\ Lett.\ B {\bf 622} (2005) 192
  hep-th/0504132,


\bibitem{Kulish}
  P.~P.~Kulish,
  {\em Twists of quantum groups and noncommutative field theory,}
  hep-th/0606056.

\bibitem{Watts}
  P.~Watts,
{\em Noncommutative string theory, the R-matrix, and Hopf algebras,}
  Phys.\ Lett.\ B {\bf 474}, 295 (2000),
  hep-th/9911026].


\bibitem{Woronowicz}
  S.~L.~Woronowicz,
 \newblock {\em Differential Calculus On Compact Matrix Pseudogroups (Quantum Groups),}
  Commun.\ Math.\ Phys.\  {\bf 122} (1989) 125.

\bibitem{AC}
  P.~Aschieri and L.~Castellani,
  \newblock {\em An Introduction to noncommutative differential geometry on quantum
  groups,}
  Int.\ J.\ Mod.\ Phys.\ A {\bf 8} (1993) 1667
  hep-th/9207084.

\bibitem{SWZ}
  P.~Schupp, P.~Watts and B.~Zumino,
  \newblock {\em Bicovariant quantum algebras and quantum Lie algebras,}
  Commun.\ Math.\ Phys.\  {\bf 157} (1993) 305
  hep-th/9210150.


\bibitem{AschieriTesi}
  P.~Aschieri,
 \newblock {\em On the geometry of inhomogeneous quantum groups,}
 math.qa/9805119, Scuola Normale Superiore di Pisa, Pubblicazioni
 Classe di Scienze, Collana Tesi.


\bibitem{NG}
P.~Aschieri et al. ~In preparation.

\bibitem{proc}
  J.~Wess,
  {\em Differential calculus and gauge transformations on a deformed space,}
  hep-th/0607251.\\
F.~Meyer, \newblock {\em Noncommutative spaces and Gravity},
Lecture given at the first Modave Summer School in Mathematical
Physics, June 2005, Modave (Belgium), hep-th/0510188. 


\bibitem{Sweedler}
M.E.~Sweedler, 
\newblock {\em Hopf Algebras}, Benjamin, New Yourk (1969),\\
E.~Abe,
\newblock {\em Hopf Algebras},
\newblock Cambridge University Press, Cambridge (1980)\\
S.~Majid,
\newblock {\em Foundations of quantum group theory},
\newblock Cambridge: University Press (1995) 606 p.


\bibitem{MadoreFC}
  M.~Buric, T.~Grammatikopoulos, J.~Madore and G.~Zoupanos,
{\em Gravity and the structure of noncommutative algebras,}
  JHEP {\bf 0604} (2006) 054,
  hep-th/0603044.



\end{thebibliography}
\end{document}